\documentclass[aps,prd,showpacs,nofootinbib,10pt,superscriptaddress,twocolumn,amsmath,amssymb,floatfix,showkeys]{revtex4-1}
\usepackage{color,amsmath,amssymb,graphicx,latexsym,lineno}
\usepackage{threeparttable,multirow,txfonts,booktabs}
\usepackage{accents,slashed,ulem,subfig}
\usepackage[colorlinks,linkcolor=blue,anchorcolor=blue,citecolor=blue,urlcolor=blue]{hyperref}
 \usepackage{float}
 \usepackage{flushend}
 \usepackage{balance}

\newlength{\dhatheight}
\newcommand{\doublehat}[1]
{%
    \settoheight{\dhatheight}{\ensuremath{\hat{#1}}}%
    \addtolength{\dhatheight}{-0.15ex}%
    \hat{\vphantom{\rule{1pt}{\dhatheight}}%
    \smash{\hat{#1}}}}
\graphicspath{{figs/}}

\begin{document}
\title{Exploring Axion-Like Particle from observation of FSRQ Ton 599 by Fermi-LAT}

\author{Jun Li$^*$}
\affiliation{School of Physical Science and Technology, Xinjiang University, Urumqi 830017, China}

\author{Xiao-Jun Bi$^\dag$}
\affiliation{Key Laboratory of Particle Astrophysics, Institute of High Energy Physics,
Chinese Academy of Sciences, Beijing 100049, China}
\affiliation{School of Physical Sciences, University of Chinese Academy of Sciences, Beijing, China}

\author{Lin-Qing Gao$^\ddag$}
\affiliation{School of Physics and Electronic Sciences, Changsha University of Science and Technology, Changsha 410114, China}

\author{Peng-Fei Yin$^\S$}
\affiliation{Key Laboratory of Particle Astrophysics, Institute of High Energy Physics,
Chinese Academy of Sciences, Beijing 100049, China}

\begin{abstract}
High energy photons traveling through astrophysical magnetic fields have the potential to undergo oscillations with axion-like particles (ALPs), resulting in modifications to the observed photon spectrum. High energy $\gamma-$ray sources with significant magnetic field strengths provide an ideal setting to investigate this phenomenon. Ton 599, a flat spectrum radio quasar with a magnetic field strength on the order of Gauss in its emission region, presents a promising opportunity for studying ALP-photon oscillations. In this study, we analyze the effects of ALP-photon oscillations on the  $\gamma$-ray spectrum of Ton 599 as observed by Fermi-LAT. Our investigation considers the potential influences of the broad-line region and dusty torus on the $\gamma-$ray spectrum of Ton 599. We set the constraints on the ALP parameters at the $95\%$ confidence level, and find that the constraints on \(g_{a\gamma}\) can reach approximately \(2 \times 10^{-12}~\mathrm{GeV}^{-1}\) for \(m_a \sim 10^{9}~\mathrm{eV}\).
\end{abstract}

\maketitle

\section{introduction}

Axion-Like Particles (ALPs), a class of extremely light pseudoscalar bosons, are anticipated in various extensions of the Standard Model \cite{Svrcek:2006yi, Arvanitaki:2009fg, Marsh:2015xka}. Unlike the axions associated with the solution to the strong CP problem within the standard model \cite{Peccei:1977hh,Peccei:1977ur,Weinberg:1977ma, Wilczek:1977pj}, ALPs offer a broader parameter space that is yet to be fully investigated. The effective coupling between ALPs and photons can induce ALP-photon oscillations in the presence of external magnetic fields. Given the prevalence of astrophysical magnetic fields, the phenomena of ALP-photon oscillation have attracted significant attention in the field of astrophysics   \cite{Raffelt:1987im,DeAngelis:2007wiw,DeAngelis:2007dqd, Hooper:2007bq, Simet:2007sa, Mirizzi:2007hr, Mirizzi:2009aj, Belikov:2010ma, DeAngelis:2011id, Horns:2012kw, HESS:2013udx, Meyer:2013pny, Tavecchio:2014yoa, Meyer:2014epa, Meyer:2014gta, Fermi-LAT:2016nkz, Meyer:2016wrm, Berenji:2016jji, Galanti:2018upl, Galanti:2018myb, Zhang:2018wpc, Liang:2018mqm, 
Xia:2018xbt,
bi2021axion, Guo:2020kiq, Li:2020pcn, Liang:2020roo, Gao:2023dvn,Gao:2023und, Majumdar:2018sbv,Xia:2019yud,eckner2022first, galanti2023alp, Dessert:2022yqq,Safdi:2018oeu}. The oscillation between ALP and photon may lead to irregularities in high energy $\gamma$-ray spectra. Various astrophysical sources, including blazars \cite{Li:2020pcn,Galanti:2018upl,Zhang:2018wpc,Gao:2023dvn,Zhou:2025txk}, GRB \cite{Gao:2023und,Troitsky:2022xso,Berezhiani:1999qh,Wang:2023okw,kamenetskaia2024polarization}, galaxy clusters \cite{Wouters:2013hua,MAGIC:2024arq,Malyshev:2018rsh}, supernova remnants \cite{bi2021axion,Li:2024ivs}, and pulsars \cite{Lloyd:2019rxg,Noordhuis:2022ljw,Lozanov:2023rcd}, have been utilized to prob the properties of ALPs. Notably, extragalactic sources, from which high energy photons traverse magnetic field environments on larger spatial scales compared to Galactic sources, hold great promise for investigating the effects of ALPs.

Among extragalactic sources, blazars have garnered significant attention for their prominence in the extragalactic $\gamma$-ray sky. Blazars are a subclass of active galactic nuclei (AGNs) characterized by the presence of relativistic jets typically oriented toward Earth. They are further classified into two categories: BL Lacertae (BL Lac) objects and flat-spectrum radio quasars (FSRQs), based on the rest-frame equivalent width of the emission lines observed in their optical spectra. FSRQs show strong broad emission lines with equivalent widths exceeding 5 $\mathring{A}$, while BL Lacs exhibit either absent or weak emission lines. The presence of these broad emission lines suggests that FSRQs contain rapidly moving gas clouds near the central black hole, known as the broad-line region (BLR). Furthermore, infrared observations of FSRQs indicate the presence of a dusty torus (DT) located beyond the BLR. In FSRQs, both the BLR and DT play critical roles in reprocessing photons from the accretion disk, resulting in the emission of low-energy radiation. The BLR primarily emits a high density of ultraviolet photons, while the DT generates infrared photons. High-energy photons emitted from the emission region can interact with these ultraviolet and infrared photons, potentially being absorbed through the $e^+e^-$ pair production process. This intricate astrophysical environment of FSRQs surpasses that of BL Lac objects, prompting previous investigations to primarily focus on BL Lac sources.

Despite the increased complexity of FSRQs compared to BL Lacs, they can serve as important targets for investigating ALP-photon oscillations. The detection of very high energy photons from FSRQs may provide evidence of  ALP-photon oscillations \cite{Tavecchio:2012um,Mena:2013baa}, offering a potential solution to avoiding the substantial absorption effects induced by BLR and DT. The analysis of the $\gamma$-ray spectra of FSRQs can also be used to set constraints on the ALP parameters \cite{Tavecchio:2014yoa,Galanti:2019sya,Galanti:2020uwd,Li:2022jgi,Pant:2023omy}. The rate of ALP-photon oscillations is significantly influenced by the intensity of external magnetic fields. The magnetic field strength within the emission region of FSRQs typically falls in the range of $\mathcal{O}(1)$ to $\mathcal{O}(10)$ G, while the magnetic field strength of many BL Lacs typically ranges from $\mathcal{O}(0.1)$ to $\mathcal{O}(1)$ G \cite{Zhang:2011sf,Tavecchio:2009zb,Tavecchio:2014yoa}. In the Third LAT AGN Catalogs, the median magnetic field strength values of FSRQs are nearly an order of magnitude larger than those of BL Lacs, as shown in Ref.~\cite{Chen:2018wck}. Therefore, FSRQs have the potential to exhibit a significant rate of ALP-photon oscillation.

In this study, we consider the observations of FSRQ Ton 599.  This source is positioned at Right Ascension (RA) = 179.88$^{\circ}$ and Declination (Dec) = 29.25$^{\circ}$, with a redshift of z=0.725. Ton 599  has been scrutinized by various instruments including EGRET \cite{thompson1995second}, Fermi-LAT \cite{Abdo:2010ge},  and VERITAS \cite{mukherjee2017veritas}. Ton 599 presents significant variability in both optical and $\gamma$-ray energy bands, with observations of this source typically classified into quiescent and bursting phases. Recently, in Ref. \cite{rajput2024investigation}, a detailed analysis of data from five distinct periods encompassing both quiescent and bursting phases was conducted using observations from Fermi-LAT. The multi-wavelength analysis indicates a magnetic field strength in the emission region on the order of Gauss. In this work, we set constraints on the ALP parameters based on these results of Ton 599.

This paper is organized as follows. 
In Section \ref{ALPPhotonOscillation}, we introduce the ALP-photon oscillation effect. In Section \ref{environment}, we introduce the astronomical environment of Ton 599 and calculate the survoval probability of photons from the FSRQ. In Section \ref{Method}, we describe the process of fitting the gamma-ray spectra and the statistical method. In Section \ref{Results}, we present the constraints on the ALP parameters from the Fermi-LAT observations of FSRQ Ton 599 and the combined constraint from the observations of all epochs. 
Finally, we conclude in Section \ref{conclusion}.

\section{ALP-photon oscillation}
\label{ALPPhotonOscillation}

The two key parameters in the effective theory influencing the ALP-photon oscillation are the ALP mass $m_a$ and the ALP-photon coupling $g_{a\gamma}$.
The state of the ALP-photon system can be characterized by the density matrix $\rho \equiv \Psi \otimes \Psi^\dagger$, where ${\Psi}\equiv {(A_\perp,~A_\parallel,~a)}^T$. Here, $a$ represents the ALP field, and $A_\perp$ and $A_\parallel$ denote the photon polarization amplitudes perpendicular and parallel to the transverse component of the external magnetic field $B_t$, respectively. When the system with energy $E \gg m_a$ traverses a homogeneous magnetic field, the density matrix $\rho$ obeys a von Neumann-like commutator equation \cite{DeAngelis:2007dqd,Mirizzi:2009aj} expressed as 
\begin{equation}\label{equ:von Neumannn-like}
    i\frac{{\rm d}\rho}{{\rm d}z} = [\rho, \mathcal{M}_0], 
\end{equation}
where $z$ denotes the distance along the propagation direction, and $\mathcal{M}_0$ is the mixing matrix encompassing various ALP and electromagnetic effects,
encompassing various ALP and electromagnetic effects $\mathcal{M}_0$ can be given by
\begin{equation}\label{equ:mixingMatrix}
    \mathcal{M}_0 = \begin{bmatrix}
\Delta_{\perp} & 0 & 0 \\
0 & \Delta_{\parallel} & \Delta_{a\gamma} \\
0 & \Delta_{a\gamma}  & \Delta_{a} 
\end{bmatrix},
\end{equation}
where $\Delta_{\perp} = \Delta_{pl} +2\Delta_{QED}$, $\Delta_{\parallel} = \Delta_{pl} + 7/2\Delta_{QED}$, $\Delta_{a}=-m_a^2/(2E)$, and $\Delta_{a\gamma}=g_{a\gamma}B_t/2$. Here, the diagonal element $\Delta_{pl} = -\omega_{pl}/(2E)$ represents the photon propagation effect in the plasma with the typical frequency $\omega_{pl}$ depending on the electron number density. The term $\Delta_{QED}=\alpha E/(45\pi)(B_{\perp}/B_{cr})^2$ describes the QED vacuum polarization effect, where $\alpha$ is the fine structure constant, and $B_{cr}=m_e^2/|e|$ denotes a critical magnetic field strength, with $m_e$ being the electron mass. The off-diagonal element $\Delta_{a\gamma}=g_{a\gamma}B_t/2$ represents the ALP-photon mixing effect.

High energy photons, emitted from extragalactic sources, pass through various astrophysical magnetic fields on their way to Earth. The entire path can be segmented into multiple segments, with the magnetic field in each segment assumed to be constant. By solving Eq. \ref{equ:von Neumannn-like}, the survival probability of the photon can be expressed as \cite{Raffelt:1987im,Mirizzi:2007hr}: \begin{equation}\label{equ:Pgaga}
P_{\gamma\gamma} = \mathrm{Tr}\left((\rho_{11}+\rho_{22})\mathcal{T}(\boldsymbol{z}) \rho(0) \mathcal{T}^{\dagger}(\boldsymbol{z})\right), 
\end{equation}
where $\mathcal{T}(\boldsymbol{z}) \equiv \prod \limits_i^n \mathcal{T}_i (\boldsymbol{z})$, and $\mathcal{T}_i(\boldsymbol{z})$ is the transferring matrix obtained from the $i$-th segment. As the polarization of very high energy $\gamma$-rays  is typically unmeasurable, the $\gamma$-ray photons emitted from the source are assumed to be unpolarized,  and $\rho(0)$ is taken to be $\text{diag}(1/2,1/2,0)$ in this case.

\section{Astronomical environment}
\label{environment}

In this section, we discuss the various astrophysical environments that influence photons originating from the source Ton 599, taking into account absorption effects and ALP-photon oscillations. These environments include the BLR, DT, blazar jet, extragalactic space, and Galactic region. Specifically, we consider ALP-photon oscillations in the blazar jet and Galactic regions, as well as the absorption effects arising from background photons in the extragalactic region, and the background photons emitted from the BLR and DT.

\subsection{Broad-line region and dusty torus}

The BLR is a high-velocity gas structure surrounding the central black hole in AGN. The gas within the BLR emits broad spectral lines and typically exhibits rapid rotational motion around the black hole. On the other hand, the DT is a large-scale toroidal structure composed of dust located in the equatorial plane of the AGN. Photons originating from the accretion disk interact with both the BLR and DT. The BLR and DT reprocesses these photons, leading to the emissions of ultraviolet and infrared light, respectively.
These ultraviolet and infrared photons subsequently interact with high-energy photons emitted from the central emission zone, absorbing them and thereby altering the observed spectrum of \( \gamma \)-ray photons.

\begin{table}[H]
  \centering
  \caption{The parameters of Ton 599 taken from \cite{rajput2024investigation}.}
  \label{tab:Parameters_BLR}
  \begin{tabular}{cc}
    \hline
    \hline
    Name of the parameters & Values\\ 
    \hline

    ${L}_{\rm disk}$ & $4.5 \times 10^{45} \; {\rm erg/s}$\\

    $\xi_{\rm BLR}$ & $0.1 $\\
    $R_{\rm BLR}$ & $1.2 \times 10^{17} \; {\rm cm} $\\

    $\xi_{\rm DT}$ & $0.5 $\\
    $R_{\rm DT}$ & $5.0 \times 10^{18} \; {\rm cm} $  \\
    $T_{\rm DT}$ & $1000 \; {\rm K} $  \\
    ${R}$ & $8.0 \times 10^{15} \; {\rm cm}$\\
    $r_{\rm VHE}$ & $8.0 \times 10^{17} \; {\rm cm} $\\

    \hline
    \hline
  \end{tabular}
\end{table}

In this study, we characterize the BLR as an infinitesimally thin spherical shell located at a distance \( R_{\rm BLR} \) from the central black hole. The BLR reprocesses the disk radiation \( L_{\rm disk} \) with a fraction denoted by \( \xi_{\rm BLR} \), which is assumed to be 0.1. The emission from the BLR occurs at a single energy corresponding to the Mg II emission line. On the other hand. the DT is modeled as a ring-shaped structure located at a distance \( R_{\rm DT} \) from the central black hole. The DT reprocesses the disk radiation with a fraction of \( \xi_{\rm DT} = 0.5 \), and emits radiation at a single energy determined by its temperature \( T_{\rm DT} \). The parameter values for the BLR and DT are adopted from Ref. \cite{rajput2024investigation}, as outlined in Table \ref{tab:Parameters_BLR}. The radius of the emission region and the distance between the emission region and the central black hole, denoted as $R$ and $r_{\rm VHE}$, are also listed. It is noteworthy that the value of \( R_{\text{BLR}} \) is not provided in Ref. \cite{rajput2024investigation}. Therefore, we have estimated an average distance based on the inner and outer radii of the BLR \( R_{\text{  \( BLR_{\text{in}} \)}} \) and \( R_{\text{  \( BLR_{\text{out}} \)}} \) given in Ref. \cite{rajput2024investigation}.
In our analysis, we utilize the open-source package agnpy \cite{nigro2022agnpy} to calculate
the absorption of $\gamma$-ray by ultraviolet and infrared photons emitting from the BLR and the DT.

\subsection{Blazar jet magnetic field}

For FSRQs like Ton 599, the photons emitted from the region from the vicinity of the central black hole traverse several astrophysical magnetic fields before reaching earth.  One of the initial regions encountered is the blazar jet, where ALP-photon oscillations may occur within the Blazar jet magnetic field (BJMF). The BJMF can be typically described as having both toroidal and poloidal components. At larger distances from the central black hole, the toroidal component dominates \cite{Pudritz:2012xj, Begelman:1984mw}, while the poloidal component becomes negligible. Therefore, in this study, we focus exclusively on the toroidal component of the magnetic field. The strength of the magnetic field in the BJMF and the electron density can be characterized as  \cite{Pudritz:2012xj, Begelman:1984mw, OSullivan:2009dsx}

\begin{align}
B_{\rm jet}(r) = B_{0} \left( \frac{r}{r_{\rm VHE}} \right)^{-1}, \\
n_{\rm el}(r) = n_{0} \left( \frac{r}{r_{\rm VHE}} \right)^{-2},
\end{align}
where \( r_{\rm VHE} \) is the distance between the emission region and the central black hole, and \( B_{0} \) and \( n_{0} \) represent the magnetic field strength and electron density at the emission region, respectively. Furthermore, it is assumed that the magnetic field diminishes beyond the maximum scale of the jet, which is taken  to be 1 kpc.

\begin{table}[H]
  \centering
  \caption{The values of the parameters $B_{0}$ and $n_{0}$ during five epochs. $B_{0}$ and $n_{0}$ are the magnetic field strength and electron density at the emission site, respectively.}
  \label{tab:Parameters}
  \begin{tabular}{cccccc}
    \hline
    \hline
    Parameters & EpochA & EpochB & EpochC & EpochD & EpochE\\ 
    \hline
    $n_{0} [\times 10^3 {\rm cm^{-3}}]$ & 0.98 & 1.19 & 1.31 & 0.89 & 0.84 \\
    $B_{0} [{\rm Gauss}]$ & 1.63 & 1.51 & 1.51 & 2.05 & 1.73 \\
    \hline
    \hline
  \end{tabular}
\end{table}

In this study, we consider the observations of Ton 599 spanning five epochs within the period from MJD 54686 to MJD 60008 \cite{rajput2024investigation}.  Each epoch encompasses a 100-day duration. Among these five epochs, one is categorized as a quiescent epoch, denoted as epoch A, while the remaining four are flaring epochs, denoted as epochs B, C, D, and E. 
During epochs B, C, and E, gamma-ray flares were concurrently observed with optical flares, whereas in epoch D, the corresponding gamma-ray flares were either weak or absent. The parameter values of $B_0$ and $n_0$ for five epoches of Ton 599 obtained from multi-wavelength analyses in Ref.~\cite{rajput2024investigation} are listed in Table~\ref{tab:Parameters}. 

\subsection{Survival probability of photons}

Considering the uplimit of magnetic field strength in the extragalactic region to be \( \mathcal{O}(1)~\mathrm{nG}\), we make the assumption that the ALP-photon oscillation effect in this region can be neglected. High energy photons are expected to be absorbed by the extragalactic background light (EBL) through the \(e^+e^-\) pair production process. To model the EBL, we adopt the results presented in \cite{Dominguez:2010wxv}.

The ALP-photon oscillation may occur within the Galactic magnetic field. The Galactic magnetic field model adopted for this investigation is the Jansson and Farrar model, as detailed in \cite{jansson2012new}. Additionally, we employ the NE2001 model \cite{Cordes:2002wz} to characterize the distribution of Galactic electron density. Notes that the Galactic magnetic field comprises both a regular and a turbulent component. Given the negligible impact of the turbulent component, we focus exclusively on the regular component.

After incorporating the aforementioned effects, the photon spectrum at Earth can be expressed as
\begin{equation}
\label{equ:df/dE} \frac{{\rm d}\Phi}{{\rm d}E} = P_{\gamma\gamma}\frac{{\rm d}\Phi_{int}}{{\rm d}E}, 
\end{equation} 
where 
${\rm d}\Phi_{int}/{\rm d}E$ represents the intrinsic spectrum, and \( P_{\gamma\gamma} \) denotes the survival probability of the photon. The survival probability \( P_{\gamma\gamma} \) is calculated numerically based on Eqs. \ref{equ:von Neumannn-like} and \ref{equ:Pgaga}. To perform these calculations, we utilize the open-source package gammaALP \cite{Meyer:2021pbp}.

\begin{figure}[htbp]
\includegraphics[width=0.45\textwidth]{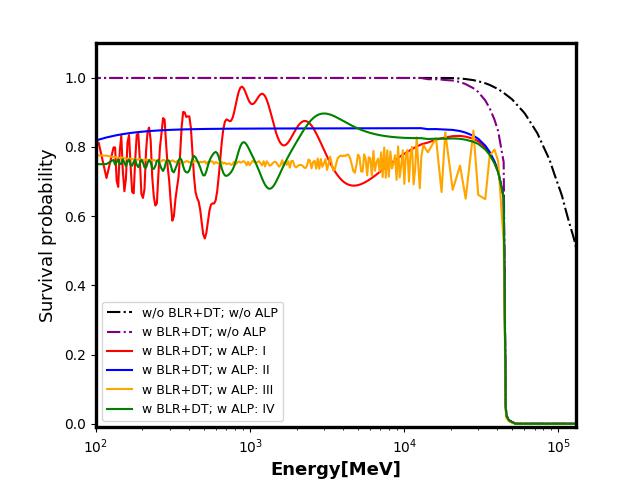}
\caption{The survival probability of photons of Ton 599 in Epoch E.}
\label{fig:surP}
\end{figure}

In Fig. \ref{fig:surP}, We illustrate the survival probability of photons in Epoch E across various scenarios. The green and purple dash-dotted lines correspond to the cases without and with the absorption in Ton 599 under the null hypothesis, respectively, in the absence of ALP effects. It is evident that the absorption effect resulting from BLR and DT can significantly suppress the $\gamma-$ray flux above $\sim 40$ GeV. We also display the survival probability of photons for four ALP parameter points, labeled as I, II, III and IV, with values of \((m_a, g_{a\gamma}) = (2\times10^{-9}~\mathrm{eV}, 5\times10^{-11}~\mathrm{GeV}^{-1})\), \( (2\times10^{-10}~\mathrm{eV}, 5\times10^{-11}~\mathrm{GeV}^{-1})\) ,  \( (2\times10^{-8}~\mathrm{eV}, 5\times10^{-11}~\mathrm{GeV}^{-1})\) and \( (2\times10^{-9}~\mathrm{eV}, 5\times 10^{-12}~\mathrm{GeV}^{-1})\), respectively. It is apparent that the conversion between the photon and ALP induces oscillatory behavior in the in the $\gamma$-ray spectrum below $\sim 10$ GeV for parameter points I and IV with $m_a\sim 10^{-9}$ eV. For parameter point II with a smaller ALP mass, the survival probability is almost energy-independent. Conversely, for parameter III with a large ALP mass, the survival probability oscillates rapidly while almost maintaining an averaged value across large energy bins. These patterns will be helpful in our understanding of the constraint results in Sec. \ref{Results}.

Our analysis indicates that the photons from Ton599 with energies above $\sim40$ GeV undergo substantial absorption by the BLR and DT. However, the MAGIC collaboration has reported the detection of photons with energies surpassing 100 GeV from Ton 599 \cite{mirzoyan2017detection}. This observed high energy $\gamma$-ray flux amounts to approximately $1.5 \times 10^{-10} \rm{ph/cm^2/s}$, 
equivalent to 0.3 Crab units. Note that this result is derived from the one-hour data collected by MAGIC on December 15, 2017, which may correspond to a period of high state for Ton 599. In contrast, our calculations in this analysis are based on observations from different epochs \cite{rajput2024investigation}, each spanning a 100-day duration. Given that the uncertainties associated with the spectral data points provided by Ref.~\cite{rajput2024investigation} are substantial above 10 GeV, the precise cutoff of the $\gamma$-ray spectrum induced by absorption effects would not significantly affect the constraints on the ALP parameters, resulting from the oscillation effects at lower energies.

\section{Method}
\label{Method}
\label{Results}
\begin{figure*}[htbp]
\includegraphics[width=0.31\textwidth, height=0.27\textwidth]{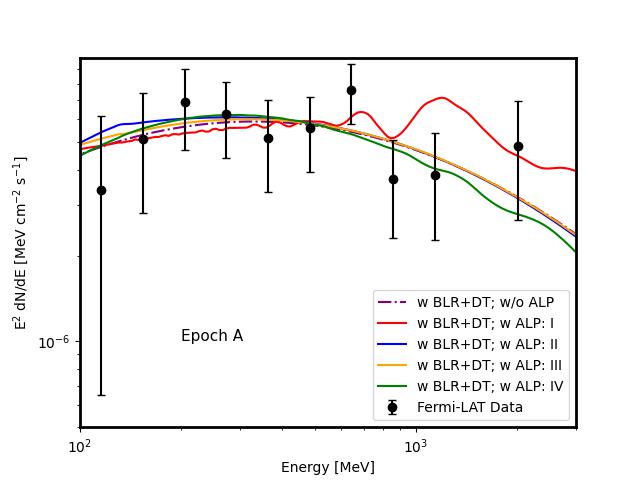}
\includegraphics[width=0.33\textwidth]{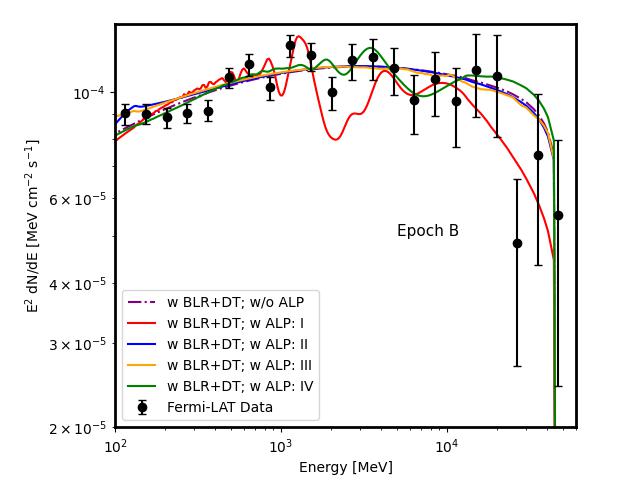}
\includegraphics[width=0.33\textwidth]{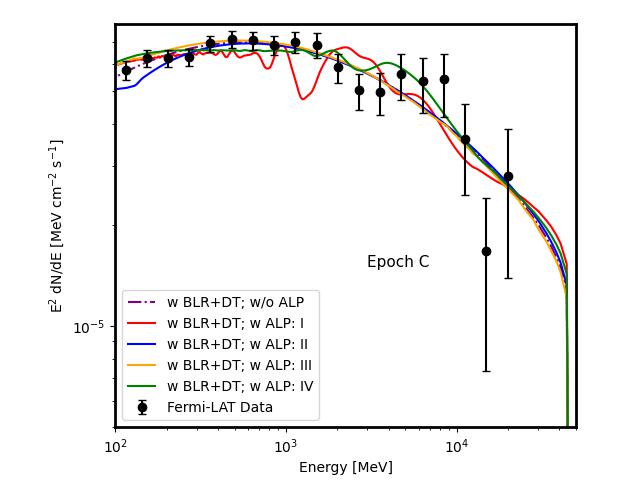}
\includegraphics[width=0.33\textwidth]{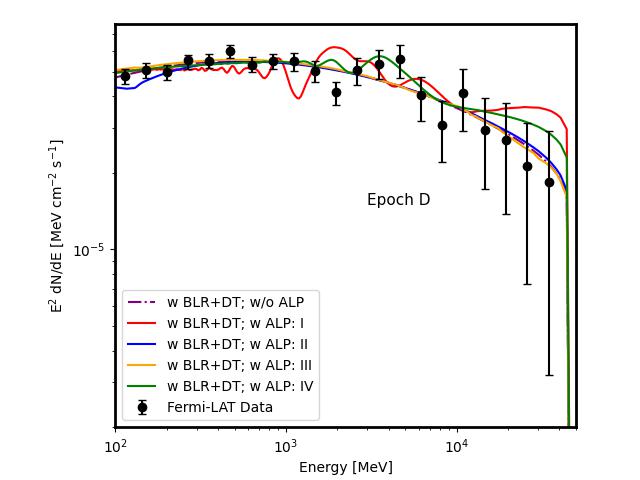}
\includegraphics[width=0.33\textwidth]{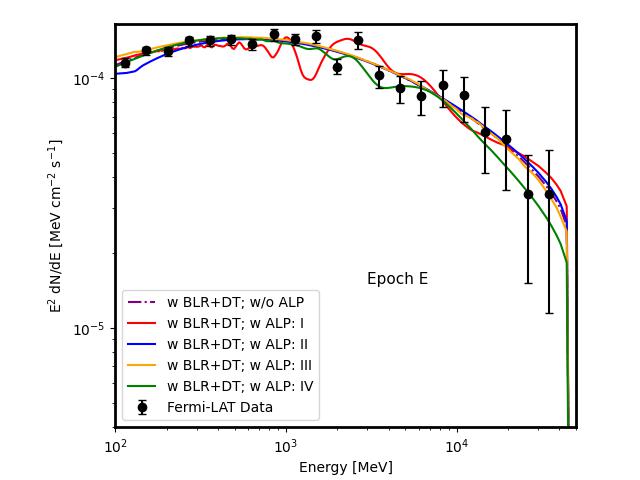}
\caption{Best-fit spectra for the five observations of Ton 599. The blue dashed lines and red solid lines represent the best-fit spectra under the null hypothesis, with and without considering the absorption effects of the BLR and DT, respectively. The yellow and green dashed lines correspond to the best-fit spectra under the ALP hypothesis, for four selected parameter points, which are the same as those in Fig.~\ref{fig:surP}.}
\label{fig:Spec}
\end{figure*}

In this section, we introduce the analysis method employed to set constraints on the ALP parameters. 
In this study, we consider the intrinsic spectrum of Ton 599 to follow a log parabola (LP) model given by
\begin{equation}
\Phi_{int}(E)=N_0 \left( \frac{E}{E_0} \right)^{- \alpha - \beta \log\left ( \frac{E}{E_0} \right )}.
\end{equation}
where $N_0$, $\alpha$ and $\beta$ are free parameters, with $E_0$ set to be 100 $\mathrm{MeV}$.

The determination of the best-fit spectrum involves minimizing the $\chi^2$ function, defined as
\begin{equation}
    \chi^2 =  \sum\limits _{j} \chi^2_{j},
\end{equation}
where $\chi^2_{j}$ denotes the $\chi^2$ function of the $j$-th epoch of Ton 599. $\chi^2_{j}$ is given by
\begin{equation}
    \chi^2_{j}= \sum\limits _{i}\frac{(\widetilde{\Phi}_i-\Phi_{i})^2}{\delta {\Phi_{i}}^2},
\end{equation}
where $\widetilde{\Phi}_i$, $\Phi_i$, and $\delta \Phi_i$ correspond to the predicted value, observed value, and experimental uncertainty of the photon flux in the $i$-th energy bin, respectively.

For a given set of ALP parameters $m_a$ and $g_{a\gamma}$, we define the test statistic (TS) as 
\begin{equation}
\label{equ:TS_CLs}
    {\rm TS}(m_a, g_{a\gamma}) = \chi^2_{\rm ALP} (\doublehat{F_0}, \doublehat{{\Gamma}}, \doublehat{{b}}; m_a, g_{a\gamma})-\chi^2_{\rm Null}(\hat{F_0}, \hat{\Gamma}, \hat{b}),
\end{equation}
where $\chi^2_{\rm Null}$ represents the best-fit $\chi^2$ value under the null hypothesis without the ALP-photon oscillation effect,  and $\chi^2_{\rm ALP}$ represents the best-fit $\chi^2$ value under the alternative hypothesis including the ALP-photon oscillation effect with the given $m_a$ and $g_{a\gamma}$. Here, the terms $(\hat{F_0}, \hat{\Gamma}, \hat{b})$ and $(\doublehat{F_0}, \doublehat{{\Gamma}}, \doublehat{{b}})$ denote the best-fit values of the intrinsic spectrum parameters under the null and alternative hypotheses, respectively.

To set constraints on ALP parameters, it is essential to understand the distribution of the TS. In cases where ALP parameters have a non-linear impact on the photon spectrum, Wilks' theorem is not applicable, and the TS distribution cannot be described by a $\chi^2$ distribution. Therefore, to derive constraints on ALP parameters, Monte Carlo simulations are required. In this study, we employ the ${\rm CL_s}$ method \cite{Junk:1999kv,Read:2002hq_cls,Lista:2016chp} to establish constraints on these parameters. This method has been used in our previous works \cite{Gao:2023dvn,Gao:2023und,Li:2024ivs,Gao:2024wpn}, and a detailed description of the method can be found in Refs.~\cite{Gao:2023dvn,Gao:2023und}.

\begin{figure*}[htbp]

\includegraphics[width=0.33\textwidth]{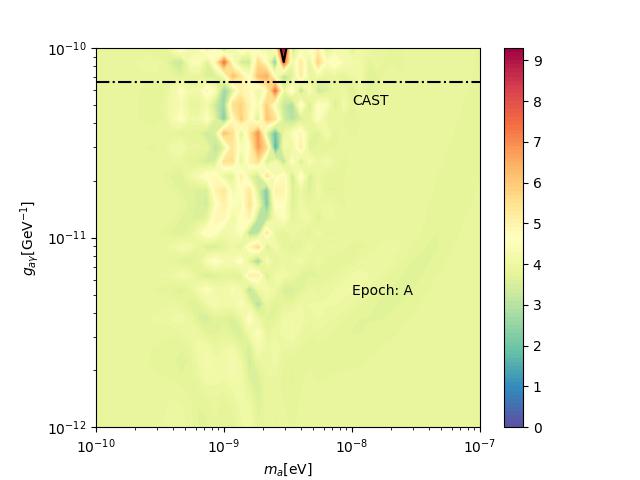}
\includegraphics[width=0.33\textwidth]{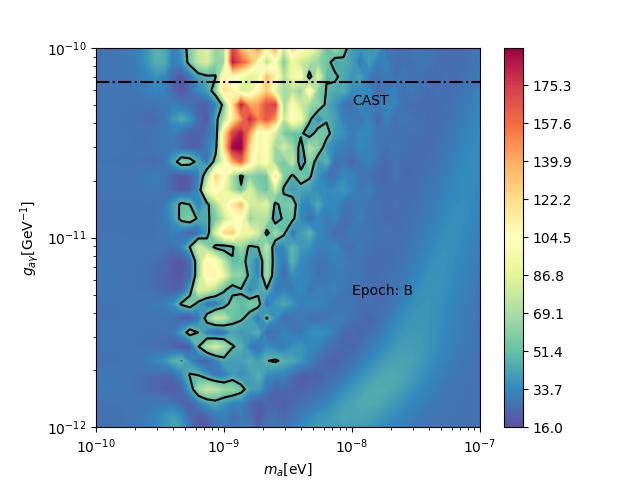}
\includegraphics[width=0.33\textwidth]{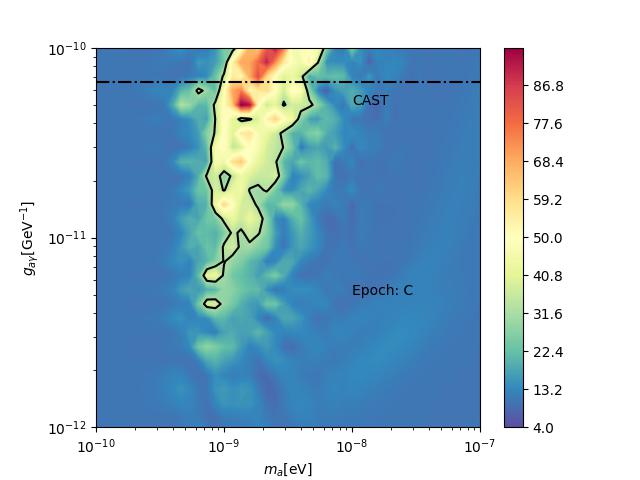}
\includegraphics[width=0.33\textwidth]{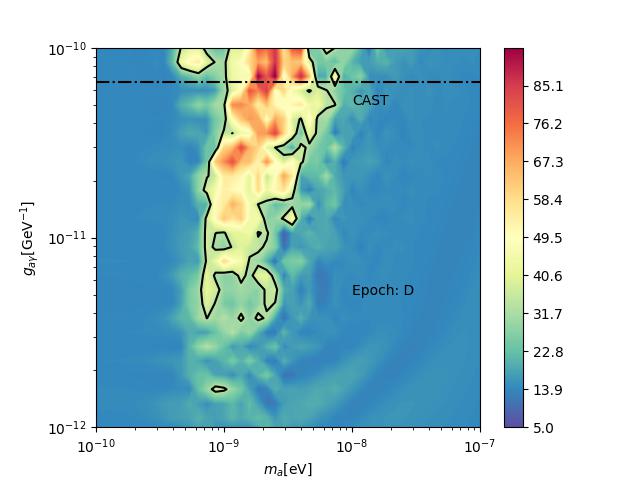}
\includegraphics[width=0.33\textwidth]{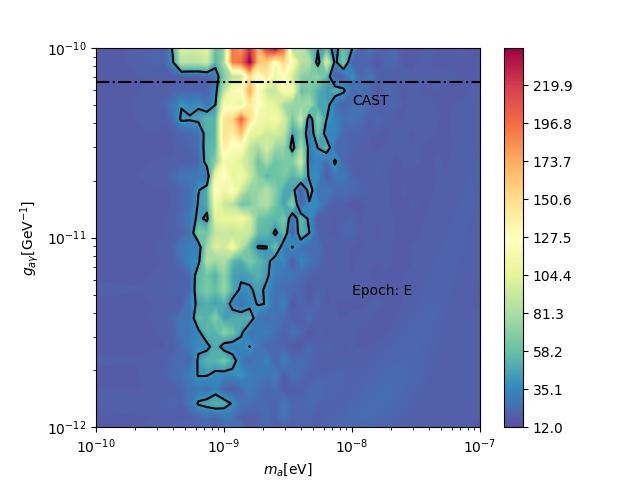}
\includegraphics[width=0.33\textwidth]{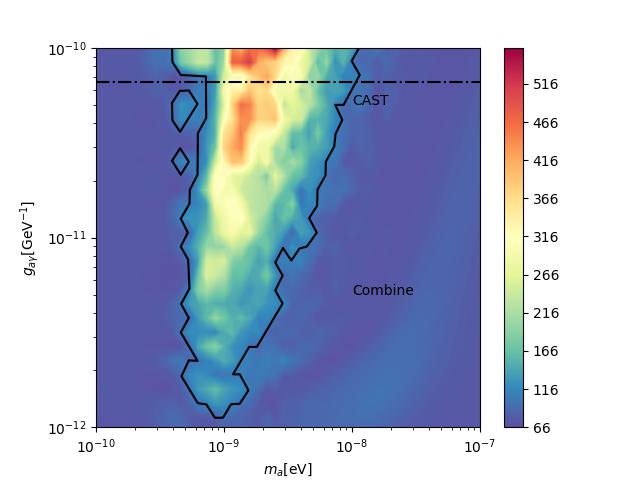}
\caption{
The TS map in the $m_a-g_{a\gamma}$ plane based on the individual and combined analyses of five epochs.
The solid black lines represent the 95\% C.L. constraints established using the $\rm CL_s$ method in this study. The dash black line denotes the constraints from the CAST experiment~\cite{CAST:2017uph}.}
\label{fig:TS}
\end{figure*}

The ${\rm CL_s}$ method can be briefly described as follows: To test whether a specific ALP parameter point $(m_a, g_{a\gamma})$ can be excluded, we first generate a set of mock data based on the expected spectrum without ALPs, denoted as $\rm \{d\}_{b}$. Next, we generate a second dataset using a similar method, but based on the energy spectrum with ALPs. The photon flux for each energy bin is drawn from a Gaussian distribution, with the expected flux as the mean and the experimental uncertainty as the standard deviation. We then calculate two TS distributions for the specific $(m_a, g_{a\gamma})$ point: one from the dataset $\rm \{d\}_{s+b}$, which includes signal and background, and one from $\rm \{d\}_b$, which contains only background. These distributions are labeled as $\{ \rm TS\}_b$ and $\{
\rm TS\}_{s+b}$, respectively. The observed TS value, denoted as ${\rm TS_{obs}}$, is calculated from the actual observed data. The ${\rm CL_s}$ value is defined as:
\[
\rm CL_s = \frac{CL_{s+b}}{CL_b},
\]
where $\rm CL_{s+b}$ and $\rm CL_b$ represent the probabilities of obtaining a TS value greater than ${\rm TS_{obs}}$ according to the distributions $\rm \{TS\}_{s+b}$ and $\rm \{TS\}_b$, respectively. If ${\rm CL_s}$ is less than 0.05, the corresponding parameter point is considered excluded at a 95\% confidence level (C.L.).

\section{Results}
\label{Results}

In this section, we present constraints on the ALP parameters derived from the observational results of the Fermi-LAT for the FSRQ Ton 599. Our analysis indicates that the observational data are consistent with the null hypothesis, assuming the absence of ALP. The best-fit spectra under the null hypothesis are shown as the blue dashed lines in Fig.~\ref{fig:Spec}, with five subfigures corresponding to Epoch A, B, C, D, and E. For comparison, we also display the best-fit spectra under the ALP hypothesis for the four parameter points, for which the survival probability of photons have been shown in Fig.~\ref{fig:surP}.

We perform a scan of the parameter space with \(m_a \in [10^{-10}, 10^{-6}]~\mathrm{eV}\) and \(g_{a\gamma} \in [10^{-13}, 10^{-10}]~\mathrm{GeV}^{-1}\), and establish constraints at the 95\% C.L. using the \(\rm CL_s\) method. The constraints derived from observations of the five epochs are shown in Fig.~\ref{fig:TS}. Due to the large flux uncertainties in the quiescent Epoch A, it cannot provide effective constraints on ALPs. The most stringent constraints are obtained from Epoch B and E. For \(m_a = 8\times10^{-10}~\mathrm{eV}\), the constraints on \(g_{a\gamma}\) approach \(1.3 \times 10^{-12}~\mathrm{GeV}^{-1}\) in Epoch B. For \(m_a = 9.2\times10^{-10}~\mathrm{eV}\), the constraint on \(g_{a\gamma}\) approach \(1.3 \times 10^{-12}~\mathrm{GeV}^{-1}\) in Epoch E. Furthermore, we present the combined constraints obtained from all observations in five epochs in Fig.~\ref{fig:TS}. These combined constraints are more stringent than those derived from any single epoch, with the best constraint on \(g_{a\gamma}\) reaching approximately \(1.1 \times 10^{-12}~\mathrm{GeV}^{-1}\) for \(m_a \sim 9 \times 10^{-10}~\mathrm{eV}\).

\begin{figure}[htbp]
\includegraphics[width=0.45\textwidth]{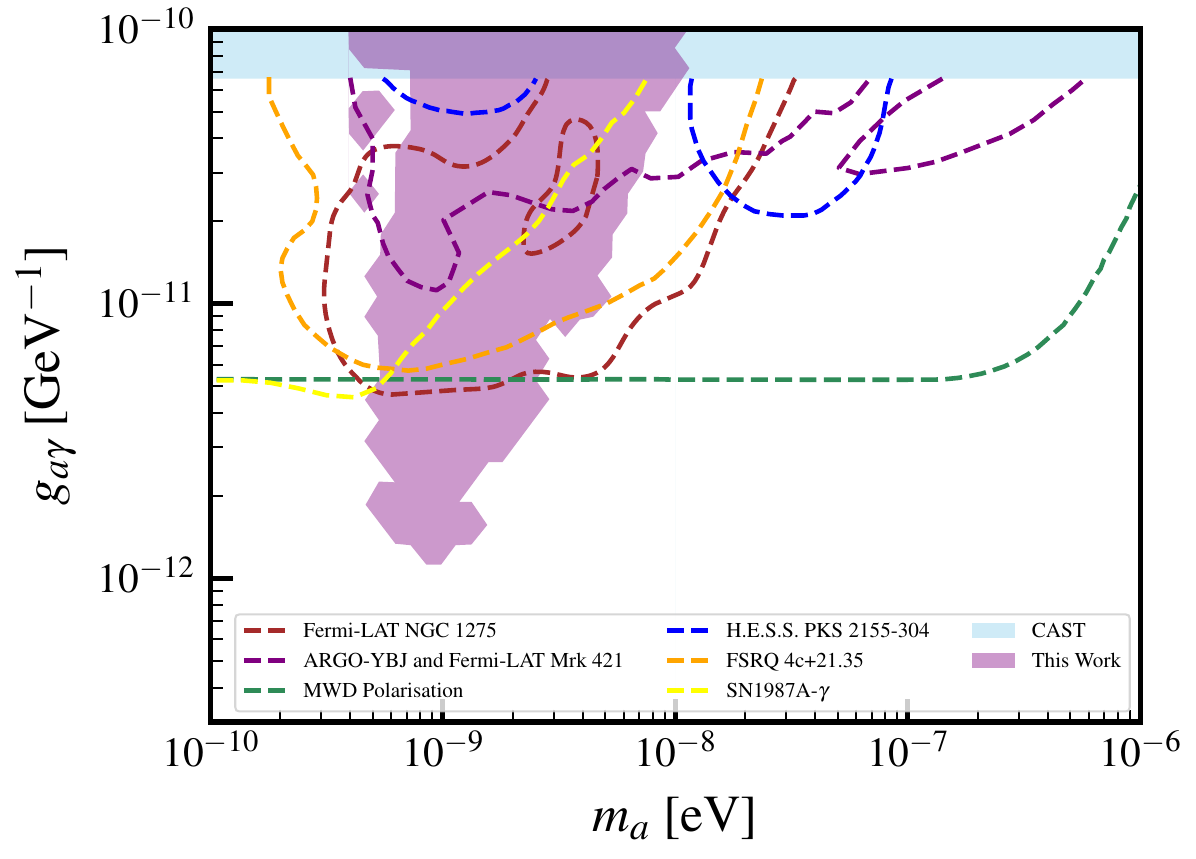}
\caption{Comparison of the constraints derived by this study with those from other studies. More constraints can be found in Ref.
\cite{AxionLimits}. 
}

\label{fig:res_comp}
\end{figure}

In our analysis,   the constraints are effectively imposed on the ALP mass at the order of $10^{-9}$ eV. This behavior can be understood as follows. Since the intrinsic spectra parameters are unknown and are treated as free parameters in the analysis, the overall attenuation of the spectra caused by the ALP would not manifest distinguishable effects in the observations, and hence would not yield constraints. The constraints arise from the oscillatory behaviors across various energies in the observed spectrum below approximately 10 GeV, which are induced by the significant energy dependent ALP-photon oscillations. As shown in Fig.~\ref{fig:surP}, the survival probability of photons exhibit significant oscillatory patterns for the two ALP parameter points with $m_a\sim10^{-9}$ eV.

The survival probability of photons can be determined by the mixing matrix denoted by Eq.~\ref{equ:mixingMatrix}, and the propagation distance, denoted as $L$. The conversion probability between the photon and ALP within a constant magnetic field can be expressed as
\begin{equation}\label{equ:probability}
P_{\gamma \rightarrow a} = \sin^2 (2 \theta) \sin^2 \left( \frac{\Delta_{\textrm{osc}}L}{2} \right),
\end{equation}
where $\theta$ represents the mixing angle given by
\begin{equation}\label{equ:mixangle}
\theta = \frac{1}{2} \arctan \left(  \frac{2\Delta_{a\gamma}}{\Delta_{\parallel} - \Delta_a } \right),
\end{equation}
and $\Delta_{\textrm{osc}}$ represents the oscillation wave number given by
\begin{equation}\label{equ:oscwavenumber}
\Delta_{\textrm{osc}} = [(\Delta_{\parallel} - \Delta_a )^2 + 4 \Delta_{a\gamma}^2 ]^{1/2}.
\end{equation}
In the scenario considered for photons with energies below $\sim \mathcal{O}(1)$ GeV, the oscillatory behaviors in the spectrum across different energies can be triggered by the mass term $\Delta_a= -m_a^2/2E$ within $\Delta_{\textrm{osc}}$.

When $m_a \sim \mathcal{O}(10^{-9})$ eV in the considered scenario, the contribution to the oscillation rate from the mass term can be comparable to the mixing term $\Delta_{a\gamma} = g_{a\gamma} B/2$ in the jet, leading to pronounced energy-dependent oscillations. As the ALP masses decrease, the energy-dependent impact of $\Delta_a$ diminishes, resulting in primarily a global attenuation in the spectrum, which does not yield significant constraints. On the other hand, for larger ALP masses, the phase term $\Delta_{\textrm{osc}} L$ would be very large, leading to extremely rapid oscillatory patterns, that may not be discerned in observations due to limited resolutions. Moreover, the substantial mass term can suppress the contribution from the mixing term $\Delta_{a\gamma}$, as shown in Eq.~\ref{equ:mixangle}, resulting in no constraints in the very high ALP mass region. These qualitative discussions are consistent with the results shown in Fig. \ref{fig:surP}, providing an understanding for the distribution of excluded regions in the parameter space. 

In Fig. \ref{fig:res_comp}, we provide a comparison of the results obtained in this study with those from other experimental studies. The shaded purple region represents the comprehensive constraint at the 95\% C.L. derived from the results of this study. The shaded blue region represents the parameter space excluded by the CAST experiment \cite{CAST:2017uph}. Additionally, constraints from various sources are included: the Fermi-LAT observation of NGC 1275 \cite{Fermi-LAT:2016nkz} (brown dashed line), the ARGO-YBJ and Fermi-LAT observations of Mrk 421 \cite{Li:2020pcn} (purple dashed line), 
the H.E.S.S. observation of PKS 2155-304 \cite{HESS:2013udx} (blue dashed line), the observation of SN1987A based on Solar Maximum Mission $\gamma-$ray data (yellow dashed line), and polarization measurements of thermal radiation from magnetic white dwarf stars \cite{Dessert:2022yqq} (green dashed line). The constraint of FSRQ 4c+21.35 from the observations of MAGIC, VERITAS, and Fermi-LAT, as reported in Ref. \cite{li2022probing}, is depicted in the orange dashed line.
Our study imposes stricter constraints compared to other works within the range  $m_a \in [4\times10^{-10},2\times10^{-9}]~\mathrm{eV}$.

\section{Conclusion}
\label{conclusion}
\flushend
In this paper, we investigate the impact of ALP-photon oscillations on the $\gamma$-ray observations of the FSRQ Ton 599.By analyzing the Fermi-LAT observations of Ton 599 in 2023, we establish constraints on the ALP parameter space at a 95\% confidence level, corresponding to a photon-ALP coupling of approximately $1.1 \times 10^{-12}$ $ {\rm GeV}^{-1}$, applicable to ALP masses within the range of $[4\times10^{-10}-2\times10^{-9}]~\mathrm{eV}$.

Our study comprehensively considers the internal and external astrophysical environments of the FSRQ Ton 599, including the BLR, DT, blazar jet, extragalactic region, and Galactic region. We have determined that the BLR and DT have minor impact on the results of spectral energy distribution fitting, leading to the conclusion that they do not influence the final outcomes significantly.

The propagation of high energy photons from FSRQ blazars through astrophysical magnetic fields on larger spatial scale provides a valuable opportunity to study ALP-photon oscillations. With the advancement of observational techniques and the development of high-precision scientific instruments, research on FSRQ blazars and ALPs will continue to deepen. Future large-scale facilities such as LHAASO, MAGIC, HESS, CTA, and DAMPE will enhance the precision of measurements and contribute to further advancements in our research efforts.

\acknowledgements
This work is supported by the National Natural Science Foundation of China under grant 12447105.

Email: $^*$mjli@xju.edu.cn
$^\dag$bixj@ihep.ac.cn
$^\ddag$gaolq@csust.edu.cn
$^\S$yinpf@ihep.ac.cn

\balance
 \bibliographystyle{apsrev}
\bibliography{Ref}
\end{document}